# A Critical Analysis of the Feasibility of Pure Strain-Actuated Giant Magnetostrictive Nanoscale Memories

P.G. Gowtham[1], G.E. Rowlands[1], and R.A. Buhrman[1]

[1]Cornell University, Ithaca, New York, 14853, USA

## Abstract

Concepts for memories based on the manipulation of giant magnetostrictive nanomagnets by stress pulses have garnered recent attention due to their potential for ultra-low energy operation in the high storage density limit. Here we discuss the feasibility of making such memories in light of the fact that the Gilbert damping of such materials is typically quite high. We report the results of numerical simulations for several classes of toggle precessional and non-toggle dissipative magnetoelastic switching modes. Material candidates for each of the several classes are analyzed and forms for the anisotropy energy density and ranges of material parameters appropriate for each material class are employed. Our study indicates that the Gilbert damping as well as the anisotropy and demagnetization energies are all crucial for determining the feasibility of magnetoelastic toggle-mode precessional switching schemes. The roles of thermal stability and thermal fluctuations for stress-pulse switching of giant magnetostrictive nanomagnets are also discussed in detail and are shown to be important in the viability, design, and footprint of magnetostrictive switching schemes.

## I. Introduction

In recent years pure electric-field based control of magnetization has become a subject of very active research. It has been demonstrated in a variety of systems ranging from multiferroic single phase materials, gated dilute ferromagnetic semiconductors [1–3], ultra-thin metallic ferromagnet/oxide interfaces [4–10] and piezoelectric/magnetoelastic composites [11–15]. Beyond the goal of establishing an understanding of the physics involved in each of these systems, this work has been strongly motivated by the fact that electrical-field based manipulation of magnetization could form the basis for a new generation of ultra-low power, non-volatile memories. Electric-field based magnetic devices are not necessarily limited by Ohmic losses during the write cycle (as can be the case in current based memories such as spin-torque magnetic random access memory (ST-MRAM)) but rather by the capacitive charging/decharging energies incurred per write cycle. As the capacitance of these devices scale with area the write energies have the potential to be as low as 1 aJ per write cycle or less.

One general approach to the electrical control of magnetism utilizes a magnetostrictive magnet/piezoelectric transducer hybrid as the active component of a nanoscale memory element. In this approach a mechanical strain is generated by an electric field within the piezoelectric substrate or film and is then transferred to a thin, nanoscale magnetostrictive magnet that is formed on top of the piezoelectric. The physical interaction driving the write cycle of these devices is the magnetoelastic interaction that describes the coupling between strain in a magnetic body and the magnetic anisotropy energy. The strain imposed upon the magnet creates an internal effective magnetic field via the magnetoelastic interaction that can exert a direct torque on the magnetization. If successfully implemented this torque can switch the magnet from one stable configuration to another, but whether imposed stresses and strains can be used to switch a magnetic element between two bi-stable states depends on the strength of the magnetoelastic coupling (or the magnetostriction). Typical values of the magnetostriction ($\lambda_s = 0.5$-$60$ ppm) in most ferromagnets yield strain and stress scales that make the process of strain-induced switching inefficient or impossible. However, considerable advances have been made in synthesizing materials both in bulk and in thin film form that have magnetostrictions that are one to two orders of magnitude larger than standard transition metal ferromagnets. These giant magnetostrictive materials allow the efficient conversion of strains into torque on the magnetization. However it is important to note that a large magnetostrictive (or magnetoelastic) effect tends to also translate into very high magnetic damping by virtue of the strong coupling between magnons and the phonon thermal bath, which has important implications, both positive and negative, for piezoelectric based magnetic devices.

In this paper we provide an analysis of the switching modes of several different implementations of piezoelectric/magnetostrictive devices. We discuss how the high damping that is generally associated with giant magnetoelasticity affects the feasibility of different approaches, and we also take other key material properties into consideration, including the saturation magnetization of the magnetostrictive element, and the form and magnitude of its magnetic anisotropy. The scope of this work excludes device concepts and physics circumscribed by magneto-elastic manipulation of domain walls in magnetic films, wires, and nanoparticle arrays [11,12,16]. Instead we focus here on analyzing various magnetoelastic reversal modes, principally within the single domain approximation, but we do extend this work to micromagnetic modeling in cases where it is not clear that the macrospin approximation provides a fully successful description of the essential physics. We enumerate potential material candidates for each of the modes evaluated and discuss the various challenges inherent in constructing reliable memory cells based on each of the reversal modes that we consider.

## II. Toggle-Mode Precessional Switching

Stress pulsing of a magnetoelastic element can be used to construct a toggle mode memory. The toggling mechanism between two stable states relies on transient dynamics of the magnetization that are initiated by an abrupt change in the anisotropy energy that is of fixed and short duration. This change in the anisotropy is created by the stress pulse and under the right conditions can generate precessional dynamics about a new effective field. This effective field can take the magnetization on a path such that when the pulse is turned off the magnetization will relax to the other stable state. This type of switching mode is referred to as toggle switching because the same sign of the stress pulse will take the magnetization from one state to the other irrespective of the initial state. We can divide the consideration of the toggle switching modes into two cases; one that utilizes a high $M_s$ in-plane magnetized element, and the other that employs perpendicular magnetic anisotropy (PMA) materials with a lower $M_s$. We make this distinction largely because of differences in the structure of the torques and stress fields required to induce a switch in these two classes of systems. The switching of in-plane giant magnetostrictive nanomagnets with sizeable out-of-plane demagnetization fields relies on the use of in-plane uniaxial stress-induced effective fields that overcome the in-plane anisotropy (~$O(10^2$ Oe)). The moment will experience a torque canting the moment out of plane and causing precession about the large demagnetization field. Thus the precessional time scales for toggling between stable in-plane states will be largely determined by the demagnetization field (and thus $M_s$). The dynamics of this mode bears striking resemblance to the dynamics in hard-axis field pulse switching of nanomagnets [17]. On the other hand, the dominant energy scale in PMA giant magnetostrictive materials is the perpendicular anisotropy energy. This energy scale can vary substantially (anywhere from $K_u \sim 10^5\text{-}10^7$ ergs/cm$^3$) depending on the materials utilized and the details of their growth. The anisotropy energy scale in these materials can be tuned into a region where stress-induced anisotropy energies can be comparable to it. A biaxial stress-induced anisotropy energy, in this geometry, can induce switching by cancelling and/or overcoming the perpendicular anisotropy energy. As we shall see, this fact and the low $M_s$ of these systems imply dynamical time scales that are substantially different from the case where in-plane magnetized materials are employed.

## A. In-Plane Magnetized Magnetostrictive Materials

We first treat the macrospin switching dynamics of an in-plane magnetized magnetostrictive nanomagnet with uniaxial anisotropy under a simple rectangular uniaxial stress pulse. Giant magnetostriction in in-plane magnetized systems have been demonstrated for sputtered polycrystalline Tb$_{0.3}$Dy$_{0.7}$Fe$_2$ (Terfenol-D) [18], and more recently in quenched Co$_x$Fe$_{1-x}$ thin film systems [19]. We assume that the uniaxial anisotropy is defined completely by the shape anisotropy of the elliptical element and that any magneto-crystalline anisotropy in the film is considerably weaker. This is a reasonable assumption for the materials considered here in the limit where the grain size is considerably smaller than the nanomagnet's dimensions. The stress field is applied by voltage pulsing an anisotropic piezoelectric film that is in contact with the nanomagnet. The proper choice of the film orientation of a piezoelectric material such as <110> lead magnesium niobate-lead titanate(PMN-PT) can ensure that an effective uniaxial in-plane strain develops along a particular crystalline axis after poling the piezo in the z-direction. We assume that the nanomagnet major axis lies along such a crystalline direction (the <110>-direction of PMN-PT) so that the shape anisotropy is coincident with the strain axis (see Figure 1 for the relevant geometry). For the analysis below we use material values appropriate to sputtered, nanocrystalline Tb$_{0.3}$Dy$_{0.7}$Fe$_2$ [18] ($M_s$ = 600 emu/cm$^3$, $\lambda_s$ = 670 ppm is the saturation magnetostriction). Nanocrystalline Tb$_{0.3}$Dy$_{0.7}$Fe$_2$ films, with a mean crystalline grain diameter $d_{grain}$ < 10 nm, can have an extremely high magnetostriction while being relatively magnetically soft with coercive fields, $H_c$ ~ 50-100 Oe, results which can be achieved by thermal processing during sputter growth at T ~ 375 °C [20]. The nanomagnet dimensions were assumed to be 80 nm (minor axis) × 135 nm (major axis) × 5 nm (thickness) yielding a shape anisotropy field $H_k = 4\pi(N_y - N_x)M_s$ = 323 Oe and $H_{demag} = 4\pi(N_z - N_y)M_s$ = 5.97 kOe. We use demagnetization factors that are correct for an elliptical cylinder [21].

The value of the Gilbert damping parameter $\alpha$ for the magnetostrictive element is quite important in determining its dynamical behavior during in-plane stress-induced toggle switching. Previous simulation results [22–24] used a value ($\alpha = 0.1$ for Terfenol-D) that, at least arguably, is considerably lower than is reasonable since that value was extracted from spin pumping in a Ni (2 nm) /Dy(5 nm) bilayer [25]. However, that bilayer material is not a good surrogate for a rare-earth transition-metal alloy (especially for $L \ne 0$ rare earth ions). In the latter case the loss contribution from direct magnon to short wavelength phonon conversion is important, as has been directly confirmed by studies of $L \ne 0$ rare earth ion doping into transition metals [26,27]. For example in-plane magnetized nanocrystalline 10% Tb-doped Py shows $\alpha$ ~ 0.8 when magnetron sputtered at 5 mtorr Ar pressure, even though the magnetostriction is small within this region of Tb doping [27]. We contend that a substantial increase in the magnetoelastic interaction in alloys with higher Tb content is likely to make $\alpha$ even larger. Magnetization rotation in a highly magnetostrictive magnet will efficiently generate longer wavelength acoustic phonons as well and heat loss will be generated when these phonons thermalize. Unfortunately, measurements of the magnetic damping parameter in polycrystalline $Tb_{0.3}Dy_{0.7}Fe_2$ do not appear to be available in the literature. However, some results on the amorphous $Tb_x[FeCo]_{1-x}$ system, achieved by using recent ultra-fast demagnetization techniques, have extracted $\alpha \sim 0.5$ for compositions (x ~ 0.3) that have high magnetostriction [28]. We can also estimate the scale for the Gilbert damping by using a formalism that takes into account direct magnon to long wavelength phonon conversion via the magnetoelastic interaction and subsequent phonon relaxation to the thermal phonon bath[29]. The damping can be estimated by the following formula:

$$\alpha = \frac{36\rho\gamma}{M_s\tau}\lambda_s^2 \left( \frac{1}{\left(\frac{c_T M_s}{2\gamma_{eff} A_{ex}}\right)^2} + \frac{1}{\left(\frac{c_L M_s}{2\gamma_{eff} A_{ex}}\right)^2} \right) \quad (1)$$

Using $M_s = 600$ emu/cm$^3$, the exchange stiffness $A_{ex} = 0.7 \times 10^{-6}$ erg/cm, a mass density $\rho = 8.5$ g/cm$^3$, Young's modulus of 65 GPa [30], Poisson ratio $\nu = 0.3$, and an acoustic damping time $\tau = 0.18$ ps [29] the result is an estimate of $\alpha \sim 1$. Given the uncertainties in the various parameters determining the Gilbert damping, we examine the magnetization dynamics for values of $\alpha$ ranging from 0.3 to 1.0.

We simulate the switching dynamics of the magnetic moment of a Terfenol-D nanomagnet at T=300 K using the Landau-Lifshitz-Gilbert form of the equation describing the precession of a magnetic moment **m**:

$$\frac{d\mathbf{m}}{dt} = -\gamma_{eff}\mathbf{m}\times\mathbf{H}_{eff}(t) - \gamma_{eff}\mathbf{m}\times\mathbf{H}_{Langevin}(t) + \alpha\mathbf{m}\times\frac{d\mathbf{m}}{dt} \qquad (2)$$

where $\gamma_{eff}$ is the gyromagnetic ratio. As $Tb_{0.3}Dy_{0.7}Fe_2$ is a rare earth – transition metal (RE-TM) ferrimagnet (or more accurately a speromagnet), the gyromagnetic ratio cannot simply be assumed to be the free electron value. Instead we use the value $\gamma_{eff} = 1.78\times 10^7$ Hz/Oe as extracted from a spin wave resonance study in the $TbFe_2$ system [31] which appears appropriate since Dy and Tb are similar in magnetic moment/atom (10 $\mu_B$ and 9 $\mu_B$ respectively) and g factor ( ~4/3 and ~3/2 respectively).

The first term in Equation (2) represents the torque on the magnetization from any applied fields, the effective stress field, and any anisotropy and demagnetization fields that might be present. The third term in the LLG represents the damping torque that acts to relax the magnetization towards the direction of the effective field and hence damp out precessional dynamics. The second term is the Gaussian-distributed Langevin field that takes into account the effect thermal fluctuations on the magnetization dynamics. From the fluctuation-dissipation theorem, $H_{Langevin}^{RMS} = \sqrt{\frac{2\alpha k_B T}{\gamma_{eff} M_s V \Delta t}}$ where $\Delta t$ is the simulation time-step [32]. Thermal fluctuations are also accounted for in our modeling by assuming that the equilibrium azimuthal and polar starting angles ($\varphi_0$ and $\psi_0 = \pi/2$ respectively) have a random mean fluctuation given by equipartition as $\delta\varphi_0^{RMS} = \sqrt{\frac{k_B T}{\left.\frac{\partial^2 E}{\partial \varphi^2}\right|_{\varphi_0} V}}$ and $\delta\psi_0^{RMS} = \sqrt{\frac{k_B T}{4\pi(N_z - N_y)M_s^2 V}}$. A $H_{bias}$ of 100 Oe was

used for our simulations which creates two stable energy minima at $\varphi_0 = \arcsin\left(\frac{H_{bias}}{H_k}\right) \sim 18°$ and $\varphi_1 = 162°$ symmetric about $\varphi = \pi/2$. This non-zero starting angle ensures that $\delta\varphi_0^{RMS} \ll \varphi_0$. This field bias is essential as the initial torque from a stress pulse depends on the initial starting angle. This angular dependence generates much larger thermally-induced fluctuations in the initial torque than a hard-axis field pulse. The hard axis bias field also reduces the energy barrier between the two stable states. For $H_{bias}$ = 100 Oe the energy barrier between the two states is $E_b$ = 1.2 eV yielding a room temperature $\Delta = E_b/k_B T = 49$. This ensures the long term thermal stability required for a magnetic memory.

To incorporate the effect of a stress pulse in Equation (2) we employ a free energy form for the effective field, $\mathbf{H}_{eff}(t) \equiv -\partial E/\partial \mathbf{m}$ that expresses the effect of a stress pulse along the x-direction of our in-plane nanomagnet with a uniaxial shape anisotropy in the x-direction. The stress enters the energy as an effective in-plane anisotropy term that adds to the shape anisotropy of the magnet (first term in Equation (3) below). The sign convention here is such that $\sigma > 0$ implies a tensile stress on the x-axis while $\sigma < 0$ implies a compressive strain. We also include the possibility of a bias field applied along the hard axis in the final term in Equation (3).

$$E(m_x, m_y, m_z) = -[2\pi(N_y - N_x)M_s^2 + \frac{3}{2}\lambda_s \sigma(t)]m_x^2 \\ + 2\pi(N_z - N_y)M_s^2 m_z^2 - H_{bias} M_s m_y \tag{3}$$

The geometry that we have assumed allows only for fast compressive-stress pulse based toggle mode switching. The application of a DC compressive stress along the x-axis only reduces the magnitude of the anisotropy and changes the position of the equilibrium magnetic angles $\varphi_0$ and $\varphi_1 = 180° - \varphi_0$ while keeping the potential wells associated with these states symmetric as well. Adiabatically increasing the value of the compressive stress moves the angles toward $\varphi = \pi/2$ until $\frac{3}{2}\lambda_s \sigma(t) \geq K_u$ but obviously can never induce a magnetic switch.

Thus the magnetoelastic memory in this geometry must make use of the transient behavior of the magnetization under a stress pulse as opposed to relying on quasistatic changes to the energy landscape. A compressive stress pulse where $\frac{3}{2}\lambda_s \sigma(t) \gg K_u$ creates a sudden change in the effective field. The resultant effective field $\mathbf{H}_{eff} = \left[\frac{3\lambda_s |\sigma| - 2K_u}{M_s} m_y + H_{bias}\right] \hat{\mathbf{y}}$ points in the y-direction and causes a torque that brings the magnetization out of plane. At this point the magnetization rotates rapidly about the very large perpendicular demagnetization field $\mathbf{H}_{demag}^{\perp} = -4\pi M_s m_z \hat{\mathbf{z}}$ and if the pulse is turned off at the right time will relax down to the opposite state at $\varphi_1 = 163$. Such a switching trajectory for our simulated nanomagnet is shown in the red curve in Figure 2. This mode of switching is set by a minimum characteristic time scale $\tau_{sw} = \frac{1}{\gamma 4\pi M_s} \sim 7.5 ps$, but the precession time will in general be longer than $\tau_{sw}$ for moderate stress pulse amplitudes, $\sigma(t) > 2K_u/3\lambda_s$, as the magnetization then cants out of plane enough to see only a fraction of the maximum possible $H_{demag}$. Larger stress pulse amplitudes result in shorter pulse durations being required as the magnetization has a larger initial excursion out of plane. For pulse durations that are longer than required for a $180°$ rotation (blue and green curves in Figure 2) $\mathbf{m}$ will exhibit damped elliptical precession about $\varphi = \pi/2$. If the stress is released during the correct portion of any of these subsequent precessional cycles the magnetization should relax down to the $\varphi_1$ state [blue curve in Figure 2], but otherwise it will relax down to the original state [green curve in Figure 2].

The prospect of a practical device working reliably in the long pulse regime appears to be rather poor. The high damping of giant magnetostrictive magnets and the large field scale of the demagnetization field yield very stringent pulse timing requirements and fast damping times for equilibration to $\varphi = \pi/2$. The natural time scale for magnetization damping in the in-plane magnetized thin film case is $\tau_d = \frac{1}{2\pi\gamma M_s \alpha}$, which ranges from 50 ps down to 15 ps for $\alpha = 0.3-1$ with $M_s = 600$ emu/cm$^3$. This high damping also results in the influence of thermal noise on the magnetization dynamics being quite strong since $H_{Langevin} \propto \sqrt{\alpha}$. Thus large stress levels with extremely short pulse durations are required in order to rotate the magnetization around the $\varphi = \pi/2$ minimum within the damping time, and to keep the precession amplitude large enough that the magnetization will deterministically relax to the reversed $\varphi_1$ state. Our simulation results for polycrystalline Tb$_{0.3}$Dy$_{0.7}$Fe$_2$ show that a high stress pulse amplitude of $\sigma = -85$ MPa with a pulse duration ~ 65 ps is required if $\alpha = 0.5$ (Figure 3a). However, the pulse duration window for which the magnetization will deterministically switch is extremely small in this case (<5 ps). This is due to the fact that the precession amplitude about the $\pi/2$ minimum at this damping gets small enough that thermal fluctuations allow only a very small window for which switching is reliable. For the lowest damping that we consider reasonable to assume, $\alpha = 0.3$, reliable switching is possible between $\tau_{pulse}$ ~ 30-60 ps at $\sigma = -85$ MPa. At a larger damping $\alpha = 0.75$ we find that the switching is non-deterministic for all pulse widths as the magnetization damps too quickly; instead very high stresses, $\sigma \geq -200$ MPa are required to generate deterministic switching of the magnetization with a pulse duration window $\tau_{pulse}$ ~ 25-45 ps (Figure 3b).

Given the high value of the expected damping we have also simulated the magnetization dynamics in the Landau Lifshitz (LL) form:

$$\frac{d\mathbf{m}}{dt} = -\gamma_{LL}(1+\alpha^2)\mathbf{m} \times (\mathbf{H}_{eff}(t) + \mathbf{H}_{Langevin}(t)) + \alpha\mathbf{m} \times \frac{d\mathbf{m}}{dt} \quad (4)$$

The LL form and the LLG form are equivalent in low damping limit ($\alpha \ll 1$) but they predict different dynamics at higher damping values. Which of these norm-preserving forms for the dynamics has the right damping form is still a subject of debate [33–37]. As one increases α in the LL form the precessional speed is kept the same while the damping is assumed to affect only the rate of decay of the precession amplitude. The damping in the LLG dynamics, on the other hand, is a viscosity term and retards the precessional speed. The effect of this retardation can be seen in the LLG dynamics as the precessional cycles move to longer times as a function of increasing damping. Our simulations show that the LL form (for fixed $\alpha$) predicts higher precessional speeds than the LLG and hence an even shorter pulse duration window for which switching is deterministic than the LLG, ~12 ps for LL as opposed to ~ 30 ps for LLG (Figure 3c).

The damping clearly plays a crucial role in the stress amplitude scale and pulse duration windows for which deterministic switching is possible, regardless of the form used to describe the dynamics. Even though the magnetostriction of $Tb_{0.3}Dy_{0.7}Fe_2$ is high and the stress required to entirely overcome the anisotropy energy is only 9.6 MPa, the fast damping time scale and increased thermal noise (set by the large damping and the out-of-plane demagnetization) means that the stress-amplitude that is required to achieve deterministic toggle switching is 10-20 times larger. In addition, the pulse duration for in-plane toggling must be extremely short, with typical pulse durations of 10-50 ps with tight time windows of 20-30 ps within which the acoustic pulse must be turned off. Given ferroelectric switching rise times on the order of ~50 ps extracted from experiment[38] and considering the acoustical resonant response of the entire piezoelectric / magnetostrictive nanostructure and acoustic ringing and inertial terms in the lattice dynamics, generation of such large stresses with the strict pulse time requirements needed for switching in this mode is likely unfeasible. In addition, the stress scales required to successfully toggle switch the giant magnetostrictive nanomagnet in this geometry are nearly as high or even higher than that for transition metal ferromagnets such as Ni ($\lambda_s \sim -38$ ppm with $\alpha = 0.045$). For example, with a 70 nm × 130 nm elliptical Ni nanomagnet with a thickness of 6 nm and a hard axis bias field of 120 Oe we should obtain switching at stress values $\sigma = +95$ MPa and $\tau_{pulse} = 0.75$ ns. Therefore the use of giant magnetostrictive nanomagnets with high damping in this toggle mode scheme confers no clear advantage over the use of a more conventional transition metal ferromagnet, and in neither case does this approach appear particularly viable for technological implementation.

## B. Magneto-Elastic Materials with PMA: Toggle Mode Switching

Certain amorphous sputtered RE/TM alloy films with perpendicular magnetic anisotropy such as a-TbFe$_2$ [39–42] and a- Tb$_{0.3}$Dy$_{0.7}$Fe$_2$ [43] have properties that may make these materials feasible for use in stress-pulse toggle switching. In certain composition ranges they exhibit large magnetostriction ($\lambda_s > 270$ ppm for a-TbFe$_2$, and both $\lambda_s$ and the effective out of plane anisotropy can be tuned over fairly wide ranges by varying the process gas pressure during sputter deposition, the target atom-substrate incidence angle, and the substrate temperature.

We consider the energy of such an out-of-plane magnetostrictive material under the influence of a magnetic field $H_{bias}$ applied in the $\hat{\mathbf{x}}$ direction and a pulsed biaxial stress:

$$E(m_x, m_y, m_z) = -[K_\perp^u - 2\pi M_s^2 - \frac{3}{2}\lambda_s \sigma_{biaxial}(t)]m_z^2 - M_s H_{bias} m_x \qquad (5)$$

Such a biaxial stress could be applied to the magnet if it is part of a patterned [001]-poled PZT thin film/ferromagnet bilayer. A schematic of this device geometry is depicted in Figure 4. When $H_{bias} = 0$, it is straightforward to see the stress pulse will not result in reliable switching since, when the tensile biaxial stress is large enough, the out of plane anisotropy becomes an easy-plane anisotropy and the equator presents a zero-torque condition on the magnetization, resulting in a 50%, or random, probability of reversal when the pulse is removed. However, reliable switching is possible for $H_{bias} > 0$ since that results in a finite canting of **m** towards the x-axis. This canting is required for the same reasons a hard-axis bias field was needed for the toggle switching of an in-plane magnetized element as discussed previously. A pulsed biaxial stress field can then in principle lead to deterministic precessional toggle switching between the +z and −z energy minima. This mode of pulsed switching is analogous to voltage pulse switching in the ultra-thin CoFeB|MgO using the voltage-controlled magnetic anisotropy effect.[5,8] Previous simulation results have also discussed this class of macrospin magnetoelastic switching in the context of a Ni|Barium-Titatate multilayer[44] and a zero-field, biaxial stress-pulse induced toggle switching scheme taking advantage of micromagnetic inhomogeneities has recently appeared in the literature[45]. Here we discuss biaxial stress-pulse switching for a broad class of giant magnetostrictive PMA magnets where we argue that the monodomain limit strictly applies throughout the switching process and extend past previous macrospin modeling by systematically thinking about how pulse-timing requirements and critical write stress amplitudes are determined by the damping, the PMA strength, and $M_s$ for values reasonable for these materials.

For our simulation study of stress-pulse toggle switching of a PMA magnet, we considered a $Tb_{33}Fe_{67}$ nanomagnet with an $M_s$ = 300 emu/cm$^3$, $K_{eff}$ = 4.0×10$^5$ ergs/cm$^3$ and $\lambda_s$ = 270 ppm. To estimate the appropriate value for the damping parameter we noted that ultrafast demagnetization measurements on $Tb_{18}Fe_{82}$ have yielded $\alpha = 0.27$. This 18-82 composition lies in a region where the magnetostriction is moderate ($\lambda_s$ ~50 ppm) [43] so we assumed that the damping will be on the same order or higher for a-TbFe$_2$ due to its high magnetostriction. Therefore we ran simulations for the range of $\alpha$ = 0.3-1. For the gyromagnetic ratio we used $\gamma_{eff}$ = 1.78×10$^7$ s$^{-1}$G$^{-1}$ which is appropriate for a-TbFe$_2$ [31]. We assumed an effective exchange constant $A_{eff} = 1 \times 10^{-6} erg \cdot cm^{-1}$ [46] implying an exchange length $l_{ex}^{no-stress} = \sqrt{\frac{A_{eff}}{K_{\perp}^{eff}}} = 15.8$ nm (in the absences of an applied stress) and $l_{ex}^{pulse} = \sqrt{\frac{A_{eff}}{2\pi M_s^2}} = 13.3$ nm (assuming that the stress pulse amplitude is just enough to cancel the out of plane anisotropy). A monodomain crossover criterion of $d_c \sim \sqrt{\frac{4\pi A_{ex}}{K_u}}$ ~ 56 nm (with the pulse off) and $d_c \sim \sqrt{\frac{2A_{ex}}{M_s^2}}$ ~ 47 nm (with the pulse on) can be calculated by considering the minimum length-scale associated with supporting thermal $\lambda$/2 confined spin wave modes [47]. The important point here is that the low $M_s$ of these systems ensures that the exchange length is still fairly long even during the switching process, which suggests that the macrospin approximation should be valid for describing the switching dynamics of this system for reasonably sized nanomagnets.

We simulated a circular element with a diameter of 60 nm and a thickness of 10 nm, under an x-axis bias field, $H_{bias}$ = 500 Oe which creates an initial canting angle of 11 degrees from the vertical (z-axis). This starting angle is sufficient to enable deterministic toggle precessional switching between the +z and −z minima via biaxial stress pulsing. The assumed device geometry, anisotropy energy density and bias field corresponded to an energy barrier $E_b$ = 4.6 eV for thermally activated reversal, and hence a room temperature thermal stability factor $\Delta = 185$.

We show selected results of the macrospin simulations of stress-pulse toggle switching of this modeled TbFe$_2$ PMA nanomagnet. Typical switching trajectories are shown in Figure 5a. The switching transition can be divided into two stages (see Figure 5b): the precessional stage that occurs when the stress field is applied, during which the dynamics of the magnetization are dominated by precession about the effective field that arises from the sum of the bias field and the easy-plane anisotropy field $\frac{3\lambda_s \sigma(t) + 2K_\perp^{eff}}{M_s} m_z$, and the dissipative stage that begins when the pulse is turned off and where the large $K_\perp^{eff}$ and the large $\alpha$ result in a comparatively quick relaxation to the other energy minimum. Thus most of the switching process is spent in the precessional phase and the entire switching process is not much longer than the actual stress pulse duration. For pulse amplitudes at or not too far above the critical stress for reversal, $\sigma \approx -2K_\perp^{eff}/3\lambda_s$ the two relevant timescales for the dynamics are set approximately by the precessional period $\tau_{sw} \approx 1/\gamma H_{bias} \approx 100$ ps of the nanomagnet and the damping time $\tau_d \sim 2/\alpha\gamma H_{bias}$. Both of these timescales are much longer than the timescales set by precession and damping about the demagnetization field in the in-plane magnetized toggle switching case. The result is that even with quite high damping one can have reliable switching over much broader pulse width windows, 200-450 ps. (Figure 6a,b). The relatively large pulse duration windows within which reliable switching is possible (as compared to the in-plane toggle mode) hold for both the LL and LLG damping. However, the difference between the two forms is evident in the PMA case (Figure 6c). At fixed $\alpha$, the LLG damping predicts a larger pulse duration window than the LL damping. Also the effective viscosity implicit within the LLG equation ensures that the switching time scales are slower than in the LL case as can also be seen in Figure 6c.

An additional and important point concerns the factors that determine the critical switching amplitude. In the in-plane toggle mode switching of the previous section, it was found that the in-plane anisotropy field was not the dominant factor in determining the stress scale required to transduce a deterministic toggle switch. Instead, we found that the stress scale was almost exclusively dependent on the need to generate a high enough precession amplitude/precession speed during the switching trajectory so as to not be damped out to the temporary equilibrium at $\varphi = \pi/2$ (at least within the damping range considered). This means that the critical stress scale to transduce a deterministic switch is essentially determined by the damping. We find that the situation is fundamentally different for the PMA based toggle memories. The critical amplitude $\sigma_c$ is nearly independent of the damping from a range of $\alpha = 0.3 - 0.75$ up until $\alpha \sim 1$ where the damping is sufficiently high (i.e. damping times equaling and/or exceeding the precessional time scale) that at $\sigma = -85$ MPa the magnetization traverses too close to the minimum at $\theta = \pi/2, \varphi = 0$. The main reason for this difference between the PMA toggle based memories and the in-plane toggle based memory lies in the role that the application of stress plays in the dynamics. First, in the in-plane case, the initial elliptical amplitude and the initial out of plane excursion of the magnetization is set by the stress pulse magnitude. Therefore the stress has to be high to generate a large enough amplitude such that the damping does not take the trajectory too close to the minimum at which point Langevin fluctuations become an appreciable part of the total effective field. This is not true in the PMA case where the initial precession amplitude about the bias field is large and the effective stress scale for initiating this precession about the bias field is the full cancellation of the perpendicular anisotropy.

Since the minimum stress-pulse amplitude required to initiate a magnetic reversal in out-of-plane toggle switching scales with $K_\perp^{eff}$ in the range of damping values considered, lowering the PMA of the nanomagnet is a straightforward way to reduce the stress and write energy requirements for this type of memory cell. Such reductions can be achieved by strain engineering through the choice of substrate, base electrode and transducer layers, by the choice of deposition parameters, and/or by post-growth annealing protocols. For example growing a $TbFe_2$ film with a strong tensile biaxial strain can substantially lower $K_\perp^{eff}$. If the PMA of such a nanomagnet can be reliably reduced to $K_\perp^{eff} = 2 \times 10^5$ ergs/cm$^3$ our simulations indicate that this would result in reliable pulse toggle switching at $\sigma \sim$ -50 MPa (corresponding to a strain amplitude on the $TbFe_2$ film of less than 0.1%) with $\tau_{pulse} \approx 400$ ps, for $0.3 \leq \alpha \leq 0.75$ and $H_{bias} \sim 250$ Oe . Electrical actuation of this level of stress/strain in the sub-ns regime, while challenging, may be possible to achieve.[48] If we again assume $M_s$ =300 emu/cm$^3$, a diameter of 60 nm and a thickness of 10 nm, this low PMA nanomagnet would still have a high thermal stability with $\Delta = 92$. The challenge, of course, is to consistently and uniformly control the residual strain in the magnetostrictive layer. It is important to note that no such tailoring (short of systematically lowering the damping) can exist in the in-plane toggle mode case.

### III.    Two-State Non-Toggle Switching

So far we have discussed toggle mode switching where the same polarity strain pulse is applied to reverse the magnetization between two bi-stable states. In this case the strain pulse acts to create a temporary field around which the magnetization precesses and the pulse is timed so that the energy landscape and magnetization relax the magnetization to the new state with the termination of the pulse.  Non-toggle mode magneto-elastic switching differs fundamentally from the precessional dynamics of toggle-mode switching, being an example of dissipative magnetization dynamics where a strain pulse of one sign destabilizes the original state (A) and creates a global energy minimum for the other state (B). The energy landscape and the damping torque completely determine the trajectory of the magnetization and the magnetization effectively "rolls" down to its new global energy minimum. Reversing the sign of the strain pulse destabilizes state B and makes state A the global energy minimum – thus ensuring a switch back to state A. There are some major advantages to this class of switching for magneto-elastic memories over toggle mode memories. Precise acoustic pulse timing is no longer an issue. The switching time scales, for reasonable stress values, can range from quasi-static to nanoseconds. In addition, the large damping typical of magnetoelastic materials does not present a challenge for achieving robust switching trajectories in deterministic switching as it does in toggle-mode memories. Below we will discuss deterministic switching for magneto-elastic materials that have two different types of magnetic anisotropy.

### C. The Case of Cubic Anisotropy

We first consider magneto-elastic materials with cubic anisotropy under the influence of a uniaxial stress field pulse. There are many epitaxial Fe-based magnetostrictive materials that exhibit a dominant cubic anisotropy when magnetron-sputter grown on oriented Cu underlayers on Si or on MgO, GaAs, or PMN-PT substrates. For example, $Fe_{81}Ga_{19}$ grown on MgO [100] or on GaAs exhibit a cubic anisotropy [49–51]. Given the low cost of these Fe-based materials compared to rare-earth alloys, it is worth investigating whether such films can be used to construct a two state memory. $Fe_{81}Ga_{19}$ on MgO exhibits easy axes along <100>. In addition, epitaxial $Fe_{81}Ga_{19}$ films have been found to have a reasonably high magnetostriction $\lambda_{100}=180$ ppm making them suitable for stress induced switching. If we assume that the cubic magnetoelastic thin-film nanomagnet has circular cross section, that the stress field is applied by a transducer along the [100] direction, and that a bias field is applied at $\varphi = \frac{\pi}{4}$ degrees, the magnetic free energy is:

$$E(m_x, m_y) = K_1 m_x^2 m_y^2 + K_1 m_z (1 - m_z^2) + 2\pi (N_z - N_{\parallel}) M_s^2 m_z^2 \\ - \frac{M_s H_{bias}}{\sqrt{2}} (m_x + m_y) - \frac{3}{2} \lambda_s \sigma(t) m_x^2 \quad (6)$$

Equation (6) shows that, in the absence of a bias field, the anisotropy energy is 4-fold symmetric in the film-plane. It is rather easy to see that it is impossible to make a two-state non-toggle switching with a simple cubic anisotropy energy and uniaxial stress field along [100]. Figure 7a shows the free energy landscape described by Equation (6) without stress applied. To create a two-state deterministic magnetostrictive device, $H_{bias}$ needs to be strong enough to eradicate the energy minima at $\varphi = \pi$ and $3\pi/2$ which strictly requires that $H_{bias} \geq 0.5 K_1 / M_s$.

Finite temperature considerations can lower this minimum bias field requirement considerably. This is due to the fact that the bias field can make the lifetime to escape the energy minima in the third quadrant and fourth quadrant small and the energy barrier to return them from the energy minima in the first quadrant extremely large. We arbitrarily set this requirement for the bias field to correspond to a lifetime of 75 µs. The typical energy barriers to hop from back to the metastable minima in the third and fourth quadrant for device volumes we will consider are on the order of several eV.

The requirement for thermal stability of the two minima in the first quadrant, given a diameter $d$ and a thickness $t_{film}$ for the nanomagnet, sets an *upper* bound on $H_{bias}$ as we require $\Delta \equiv E_b / k_b T > 40$ at room temperature between the two states (see Figure 7c). It is desirable that this upper bound is high enough that there is some degree of tolerance to the value of the bias field at device dimensions that are employed. This sets requirements on the minimum volume of the cylindical nanomagnet that are dependent on $K_1$.

For a circular element with $d = 100$ nm, $t_{film} = 12.5$ nm and $K_1 = 1.5 \times 10^5$ ergs/cm$^3$, two-state non-toggle switching with the required thermal stability can only occur for $H_{bias}$ between 50 - 56 Oe. This is too small a range of acceptable bias fields. However, by increasing $t_{film}$ to 15 nm the bias field range grows to $H_{bias}$ = 50 - 90 Oe which is an acceptable range. For $K_1 = 2.0 \times 10^5$ erg/cm$^3$ with $d = 100$ nm and $t_{film} = 12.5$ nm, there is an appreciable region of bias field (~65-120 Oe) for which $E_{barrier} / k_B T > 42$. For $K_1 = 2.5 \times 10^5$ ergs/cm$^3$, the bias range goes from 90 – 190 Oe for the same volume. The main point here is that, given the scale for the cubic anisotropy in Fe$_{81}$Ga$_{19}$, careful attention must be paid to the actual values of the anisotropy constants, device lateral dimensions, film thickness, and the exchange bias strength in order to ensure device stability in the sub-100 nm diameter regime.

We now discuss the dynamics for a simulated case where $d = 100$ nm, $t_{film} = 12.5$ nm, $K_1 = 2.0 \times 10^5$ ergs/cm$^3$, $H_{bias} = 85$ Oe, and $M_s = 1300$ emu/cm$^3$. Two stable minima exist at $\varphi = 10°$ and $\varphi = 80°$. Figure 7b shows the effect of the stress pulse on the energy landscape. When a compressive stress $\sigma > \sigma_c$ is applied, the potential minimum at $\varphi = 10°$ is rendered unstable and the magnetization follows the free energy gradient to $\varphi = 80°$ (green curve). Since the stress field is applied along [100] the magnetization first switches to a minima very close to but greater than $\varphi = 80°$ and when the stress is released it gently relaxes down to the zero stress minimum at $\varphi = 80°$. In order to switch from $\varphi = 80°$ to $\varphi = 10°$ we need to reverse the sign of the applied stress field to tensile (red curve). A memory constructed on these principles is thus non-toggle.

The magnetization-switching trajectory is simple and follows the dissipative dynamics dictated by the free energy landscape (see Figure 8a). We have assumed a damping of $\alpha = 0.1$ for the Fe$_{81}$Ga$_{19}$ system, based on previous measurements[52] and as confirmed by our own. Higher damping only ends up speeding up the switching and ring-down process. Figure 8b shows the simulated stress amplitude and pulse switching probability phase diagram at room temperature.

Ultimately, we must take the macrospin estimates for device parameters as only a rough guide. The macrospin dynamics approximate the true micromagnetics less and less well as the device diameter gets larger. The main reason for this is the large $M_s$ of Fe$_{81}$Ga$_{19}$ and the tendency of the magnetization to curl at the sample edges. Accordingly we have performed T = 0 °K micromagnetic simulations in OOMMF.[53] An exchange bias field $H_{bias} = 85$ Oe was applied at $\varphi = 45°$ and we assume $K_1 = 2.0 \times 10^5$ ergs/cm$^3$, $M_s = 1300$ emu/cm$^3$, and $A_{ex} = 1.9 \times 10^{-6}$ erg/cm. Micromagnetics show that the macrospin picture quantitatively captures the switching dynamics, the angular positions of the stables states ($\varphi_0 \rightarrow \sim 10°$ and $\varphi_1 \rightarrow \sim 80°$) and the critical stress amplitude at ($\sigma \sim 30$ MPa) when the device diameter $d < 75$ nm. The switching is essentially a rigid in-plane rotation of the magnetization from $\varphi_0$ to $\varphi_1$. However, we chose to show the switching for an element with $d = 100$ nm because it allowed for thermal stability of the devices in a region of thickness ($t_{film} = 12$-$15$ nm) where $H_{bias} \sim 50$-$100$ Oe at room temperature could be reasonably expected. The initial average magnetization angle is larger ($\varphi_0 \rightarrow \sim 19°$ and $\varphi_1 \rightarrow \sim 71°$) than would be predicted by macrospin for a $d = 100$ nm element. This is due to the magnetization curling at the devices edges at $d = 100$ nm (see Figure 8c). Despite the fact that magnetization profile differs from the macrospin picture we find that there is no appreciable difference between the stress scales required for switching, or the basic switching mechanism.

The stress amplitude scale for writing the simulated Fe$_{81}$Ga$_{19}$ element at $\sim 30$ MPa is not excessively high and there are essentially no demands on the acoustic pulse width requirements. These memories can thus be written at pulse amplitudes of $\sim 30$ MPa with acoustical pulse widths of $\sim 10$ ns. These numbers do not represent a major challenge from the acoustical transduction point of view. The drawbacks to this scheme are the necessity of growing high quality single crystal thin films of Fe$_{81}$Ga$_{19}$ on a piezoelectric substrate that can generate large enough strain to switch the magnet (e.g. PMN-PT) and difficulties associated with tailoring the magnetocrystalline anisotropy $K_1$ and ensuring thermal stability at low lateral device dimensions.

### D. The Case of Uniaxial Anisotropy

Lastly we discuss deterministic (non-toggle) switching of an in-plane giant magnetostrictive magnet with uniaxial anisotropy. In-plane magnetized polycrystalline TbDyFe patterned into elliptical nanomagnets could serve as a potential candidate material in such a memory scheme. To implement deterministic switching in this geometry a bias field $H_{bias}$ is applied along the hard axis of the nanomagnet. This generates two stable minima at $\varphi_0$ and $180 - \varphi_0$ symmetric about the hard axis. The axis of the stress pulse then needs to be non-collinear with respect to the easy axis in order to break the symmetry of the potential wells and drive the transition to the selected equilibrium position. Figure 9 below shows a schematic of the situation. When a stress pulse is applied in the direction that makes an angle $\beta$ with respect to the easy axis of the nanomagnet, $0° < \beta < 90°$, the free energy within the macrospin approximation becomes:

$$E(m_x, m_y, m_z) = -[2\pi(N_y - N_x)M_s^2 m_x^2 + 2\pi(N_z - N_y)M_s^2 m_z^2 - H_{bias} M_s m_y \\ + \frac{3}{2M_s}\lambda_s \sigma(t) \cdot (\cos(\beta)m_y - \sin(\beta)m_x)^2 \quad (7)$$

From Equation (7) it can be seen that a sufficiently strong compressive stress pulse can switch the magnetization between $\varphi_0$ and $180° - \varphi_0$, but only if $\varphi_0$ is between $\beta$ and $90°$. To see why this condition is necessary, we look at the magnetization dynamics in the high stress limit when $0 < \varphi_0 < \beta$. During such a strong pulse the magnetization will see a hard axis appear at $\varphi \approx \beta$ and hence will rotate towards the new easy axis at $\varphi = \beta - 90$, but when the stress pulse is turned off the magnetization will equilibrate back to $\varphi_0$. This situation is represented by the green trajectory shown in Figure 11a.

But when $\beta < \varphi_0 < 90^\text{o}$, a sufficiently strong compressive stress pulse defines a new easy axis close to $\varphi = 90^\text{o} + \beta$ and when the pulse is turned off the magnetization will relax to $\varphi = 180 - \varphi_0$ (blue trajectory in Figure 11a). Similarly the possibility of switching from $180^\text{o} - \varphi$ to $\varphi$ with a *tensile* strain depends on whether $90^\text{o} < 180^\text{o} - \varphi < 90^\text{o} + \beta$. Thus $\beta = 45^\text{o}$ is the optimal situation as then the energy landscape becomes mirror symmetric about the hard axis and the amplitude of the required switching stress (voltage) are equal. This scheme is quite similar to the case of deterministic switching in biaxial anisotropy systems (with the coordinate system rotated by $45^\text{o}$). We note that a set of papers[54–56] have previously proposed this particular case as a candidate for non-toggle magnetoelectric memory and have experimentally demonstrated operation of such a memory in the large feature-size (i.e. extended film) limit.[55]

We argue here that in-plane giant magnetostrictive magnets operated in the non-toggle mode could be a good candidate for constructing memories with low write stress amplitude, and nanosecond-scale write time operation. However, as we will discuss, the prospects of this type of switching mode being suitable for implementation in ultrahigh density memory appear to be rather poor. The main reason for this lies in the hard axis bias field requirements for maintaining low write error rates and the effect that such a hard axis bias field will have on the long term thermal stability of the element. At $T = 0$ °K the requirement on $H_{bias}$ is only that it be strong enough that $\varphi_0 > 45°$. However, this is no longer sufficient at finite temperature where thermal fluctuations imply a thermal, Gaussian distribution of the initial orientation of the magnetization direction $\varphi_0$ about $\varphi_0$. If a significant component of this angular distribution falls below 45 degrees there will be a high write error rate. Thus we must ensure that $H_{bias}$ is high enough that the probability of $\varphi < 45°$ is extremely low. We have selected the requirement that $\varphi < 45°$ is a $8\sigma$ event where $\sigma$ is the standard deviation of $\varphi$ about $\varphi_0$ and is given by the relation

$\sigma = \sqrt{\dfrac{k_B T}{\left.\dfrac{\partial^2 E}{\partial \varphi^2}\right|_{\varphi_0} V}}$ . However, $H_{bias}$ must be low enough to be technologically feasible, but also must not exceed a value that compromises the energy barrier between the two potential minima – thus rendering the nanomagnet thermally unstable. These minimum and maximum requirements on $H_{bias}$ puts significant constraints on the minimum size of the nanomagnet that can be used in this device approach. It also sets some rather tight requirements on the hard axis bias field, as we shall see.

We first discuss the effects of these requirements in the case of a relatively large magnetostrictive device. We assume the use of a polycrystalline Tb$_{0.3}$Dy$_{0.7}$Fe$_2$ element having $M_s = 600$ emu/cm$^3$ and an elliptical cross section of 400×900 nm$^2$ and a thickness $t_{film} = 12.5$ nm. This results in a shape anisotropy field $H_k \approx 260$ Oe. We find that for an applied hard axis bias field $H_{bias} \sim 200$ Oe, a field strength that can be reasonably engineered on-chip, the equilibrium angle of the element is $\varphi_0 \approx 51°$ and its root mean square (RMS) angular fluctuation amplitude is $\delta\varphi^{RMS} \approx 0.75°$. Thus element's anisotropy field and the assumed hard axis biasing conditions just satisfy the assumed requirement that $\varphi_0 - 8\delta\varphi^{RMS} > 45°$ (see Figure 10b). The magnetic energy barrier to thermal energy ratio for the element at $H_{bias} = 200$ Oe is $\Delta = E_b / k_B T$

≈ 350, which easily satisfies the long-term thermal stability requirement (see Figure 10a), and which also provides some latitude for the use of a slightly higher $H_{bias}$ if desired to further reduce the write error rate.

It is straightforward to see from these numbers that if the area of the magnetostrictive element is substantially reduced below 400×900 nm² there must be a corresponding increase in $H_k$ and hence in $H_{bias}$ if the write error rate for the device is to remain acceptable. Of course an increase in the thickness of the element can partially reduce the increase in fluctuation amplitude due to the decrease in the magnetic area, but the feasible range of thickness variation cannot match the effect of, for example, reducing the cross-sectional area by a factor of 10 to 100, with the latter, arguably, being the minimum required for high density memory applications. While perhaps a strong shape anisotropy and an increased $t_{film}$ can yield the required $H_k \geq 1$ kOe, the fact that in this deterministic mode of magnetostrictive switching we must also have $H_{bias} \sim H_k$ results in a bias field requirement that is not technologically feasible. We could of course allow the write error rate to be much larger than indicated by an $8\sigma$ fluctuation probability, but this would only relax the requirement on $H_{bias}$ marginally, which always must be such that $\varphi_0 >$ 45°. Thus the deterministic magnetostrictive device is not a viable candidate for ultra-high density memory. Instead this approach is only feasible for devices with lateral area $\geq 10^5$ nm² .

While the requirement of a large footprint is a limitation of the deterministic magnetostrictive memory element, this device does have the significant advantage that the stress scale required to switch the memory is quite low. We have simulated T = 300 °K macrospin switching dynamics for a 400×900 nm² ellipse with thickness $t_{film}$ = 12.5 nm with $H_{bias}$ = 200 Oe such that $\varphi_0 \sim 51°$. The Gilbert damping parameter was set to $\alpha = 0.5$ and magnetostriction $\lambda_s =$ 670 ppm. The magnetization switches by simple rotation from $\varphi_0 = 51°$ to $\varphi_1 = 129°$ that is driven by the stress pulse induced change in the energy landscape (see Figure 11a). Phase diagram results are provided in Figure 11b where the switching from $\varphi_0 = 51°$ to $\varphi_1 = 129°$ shows a 100% switching probability for stresses as low as $\sigma = -5$ MPa for pulse widths as short as 1 ns.

Since the dimensions of the ellipse are large enough that the macrospin picture is not strictly valid, we have also conducted T = 0 K micromagnetic simulations of the stress-pulse induced reversal in this geometry. We find that the trajectories are essentially well described by a quasi-coherent rotation with non-uniformities in the magnetization being more pronounced at the ellipse edges (see Figure 11c). The minimum stress pulse amplitude for switching is even lower than that predicted by macrospin at $\sigma = -3$ MPa. This stress scale for switching is substantially lower than any of the switching mode schemes discussed before. Despite the fact that this scheme is not scalable down into the 100-200 nm size regime, it can be appropriate for larger footprint memories that can be written at very low write stress pulse amplitudes.

## IV. CONCLUSION

The physical properties of giant magnetostrictive magnets (particularly of the rare-earth based TbFe$_2$ and Tb$_{0.3}$Dy$_{0.7}$Fe$_2$ alloys) place severe restrictions on the viability of such materials for use in fast, ultra-high density, low energy consumption data storage. We have enumerated the various potential problems that might arise from the characteristically high damping of giant magnetostrictive nanomagnets in toggle-mode switching. We have also discussed the role that thermal fluctuations have on the various switching modes and the challenges involved in maintaining long-time device thermal stability that arise mainly from the necessity of employing hard axis bias fields.

It is clear that the task of constructing a reliable memory using pure stress induced reversal of giant magnetostrictive magnets will be, when possible, a question of trade-offs and careful engineering. PMA based giant magnetostrictive nanomagnets can be made extremely small ($d < 50$ nm) while still maintaining thermal stability. The small diameter and low cross-sectional area of these PMA giant magnetostrictive devices could, in principle, lead to very low capacitive write energies. The counterpoint is that the stress fields required to switch the device are not necessarily small and the acoustical pulse timing requirements are demanding. However, it might be possible to tune the magnetostriction $\lambda_s$, $K_\perp$, and $M_s$ (either by adjustment of the growth conditions of the magnetostrictive magnet or by engineering the RE-TM multilayers appropriately) in order to significantly reduce the pulse amplitudes required for switching (down into the 20-50 MPa range) and reduce the required in-plane bias field – without compromising thermal stability of the bit. Such tuning must be carried out carefully. As we have discussed, the Gilbert damping $\alpha$, $\lambda_s$, $K_\perp$, and $M_s$ can all affect the pure stress-driven switching process and device thermal stability in ways that are certainly interlinked and not necessarily complementary.

Two state non-toggle memories such as we described in Section III D could have extremely low stress write amplitudes and non-restrictive pulse requirements. However, the trade-off arises from thermal stability considerations and such a switching scheme is not scalable down into the 100-200 nm size regime. Despite this limitation there may well be a place for durable memories with very low write stress pulse amplitudes and low write energies that operate reliably in the nanosecond regime.

## ACKNOWLEDGEMENTS

We thank R.B. van Dover, W.E. Bailey, C. Vittoria, J.T. Heron, T. Gosavi, and S. Bhave for fruitful discussions. We also thank D.C. Ralph and T. Moriyama for comments and suggestions on the manuscript. This work was supported by the Office of Naval Research and the Army Research Office.

[34] M. Stiles, W. Saslow, M. Donahue, and A. Zangwill, Phys. Rev. B **75**, 214423 (2007).

[35] N. Smith, Phys. Rev. B **78**, 216401 (2008).

[36] S. Iida, J. Phys. Chem. Solids **24**, 625 (1963).

[37] T.L. Gilbert, IEEE Trans. Magn. **40**, 3443 (2004).

[38] J. Li, B. Nagaraj, H. Liang, W. Cao, C.H. Lee, and R. Ramesh, Appl. Phys. Lett. **84**, 1174 (2004).

[39] R.B. van Dover, M. Hong, E.M. Gyorgy, J.F. Dillon, and S.D. Albiston, J. Appl. Phys. **57**, 3897 (1985).

[40] F. Hellman, Appl. Phys. Lett. **64**, 1947 (1994).

[41] F. Hellman, R.B. van Dover, and E.M. Gyorgy, Appl. Phys. Lett. **50**, 296 (1987).

[42] T. Niihara, S. Takayama, and Y. Sugita, IEEE Trans. Magn. **21**, 1638 (1985).

[43] P.I. Williams and P.J. Grundy, J. Phys. D Appl. Phys. **27**, 897 (1994).

[44] M. Ghidini, R. Pellicelli, J.L. Prieto, X. Moya, J. Soussi, J. Briscoe, S. Dunn, and N.D. Mathur, Nat. Commun. **4**, 1453 (2013).

[45] J.-M. Hu, T. Yang, J. Wang, H. Huang, J. Zhang, L.-Q. Chen, and C.-W. Nan, Nano Lett. **15**, 616 (2015).

[46] F. Hellman, A.L. Shapiro, E.N. Abarra, R.A. Robinson, R.P. Hjelm, P.A. Seeger, J.J. Rhyne, and J.I. Suzuki, Phys. Rev. B **59**, 408 (1999).

[47] J.Z. Sun, R.P. Robertazzi, J. Nowak, P.L. Trouilloud, G. Hu, D.W. Abraham, M.C. Gaidis, S.L. Brown, E.J. O'Sullivan, W.J. Gallagher, and D.C. Worledge, Phys. Rev. B **84**, 064413 (2011).

[48] A. Grigoriev, D.H. Do, D.M. Kim, C.B. Eom, P.G. Evans, B. Adams, and E.M. Dufresne, Appl. Phys. Lett. **89**, 1 (2006).

[49] A. Butera, J. Gómez, J.L. Weston, and J.A. Barnard, J. Appl. Phys. **98**, 033901 (2005).

[50] A. Butera, J. Gómez, J.A. Barnard, and J.L. Weston, Phys. B Condens. Matter **384**, 262 (2006).

[51] A. Butera, J.L. Weston, and J.A. Barnard, J. Magn. Magn. Mater. **284**, 17 (2004).
33

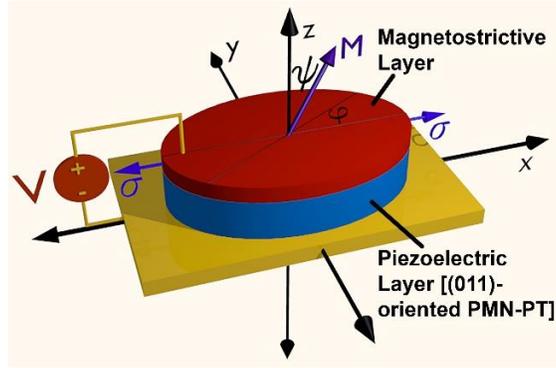

Figure 1. Magnetoelastic elliptical memory element schematic with associated coordinate system for in-plane stress-pulse induced toggle switching. Here $\mathbf{M}$ is the magnetization vector with $\psi$ and $\varphi$ being polar and azimuthal angles. For the in-plane toggle switching case, the initial normalized magnetization $\mathbf{m}_0 = \cos\varphi_0 \hat{\mathbf{x}} + \sin\varphi_0 \hat{\mathbf{y}}$ and is in the film plane with $\varphi_0 = \arcsin[H_{bias}/H_k]$ and $\mathbf{H}_{bias} = H_{bias}\hat{\mathbf{y}}$.

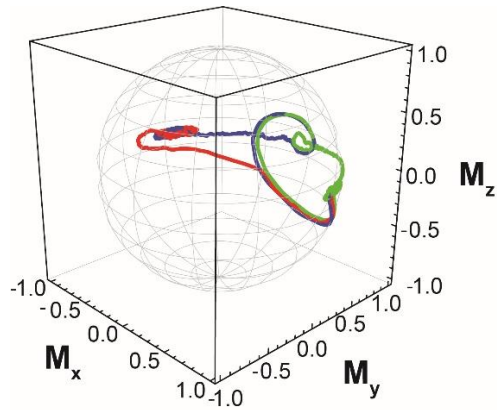

Figure 2. Toggle switching trajectory for an in-plane magnetized polycrystalline $Tb_{0.3}Dy_{0.7}Fe_2$ element with $\alpha_{LLG} = 0.3$, $\sigma = -120$ MPa, and $\tau_{pulse} = 50$ ps (red) and 125 ps (blue) and 160 ps (green).

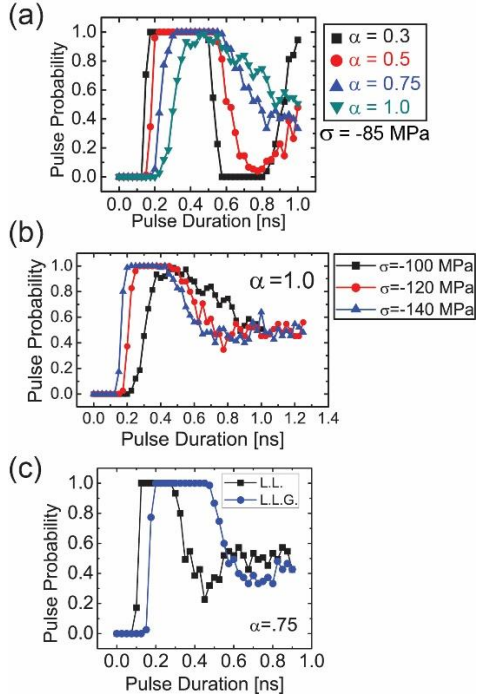

Figure 3. a) Effect of the Gilbert damping on pulse switching probability statistics for $\sigma = -85$ MPa. b) Effect of increasing stress pulse amplitude for high damping $\alpha_{LLG} = 0.75$. Very high stress pulses ( >200 MPa) are required to allow precession to be fast enough to cause a switch before dynamics are damped out. c) Comparison of switching statistics for the LL and LLG dynamics at $\sigma = -200$ MPa, $\alpha = 0.75$. The LL dynamics exhibits faster precession than the LLG for a given torque implying shorter windows of reliability and requirements for faster pulses.

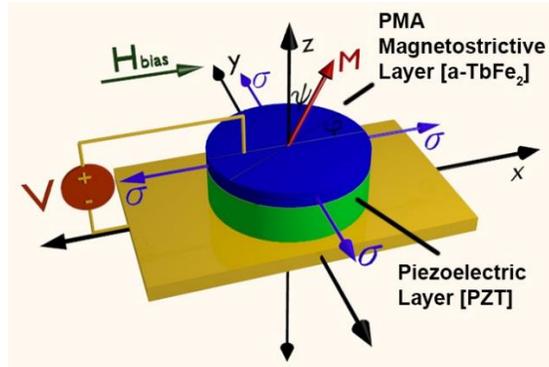

Figure 4. Schematic of TbFe$_2$ magnetic element under biaxial stress generated by a PZT layer. Here the initial normalized magnetization $\mathbf{m}_0 = \cos\psi_0 \hat{\mathbf{z}} + \sin\psi_0 \hat{\mathbf{x}}$ is predominantly out of the film plane with a cant $\psi_0 = \arcsin[H_{bias}/H_k]$ in the x-direction provided by $\mathbf{H}_{bias} = H_{bias}\hat{\mathbf{x}}$.

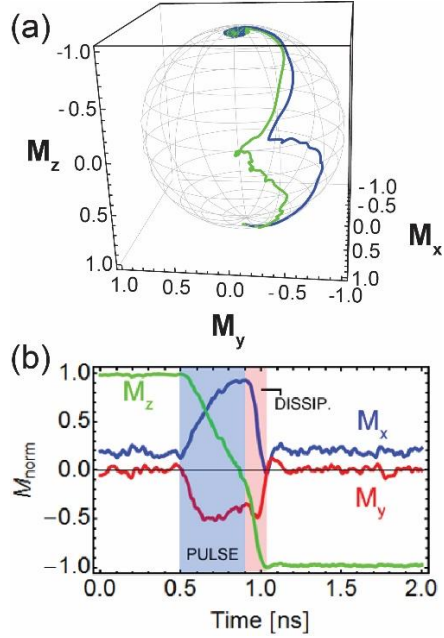

Figure 5. a) Switching trajectories for a TbFe$_2$ nanomagnet under a pulsed biaxial stress $\sigma = -85$ MPa, $\tau_{pulse} = 400$ ps (green) and $\sigma = -120$ MPa and $\tau_{pulse} = 300$ ps (blue) b) Switching trajectory time trace for {m$_x$,m$_y$,m$_z$} for $\sigma = -85$ MPa. The pulse is initiated at t = 500 ps. The blue region denotes when precession about $H_{bias}$ dominates (i.e. while the pulse is on) and the red when the dissipative dynamics rapidly damp the system down to the other equilibrium point.

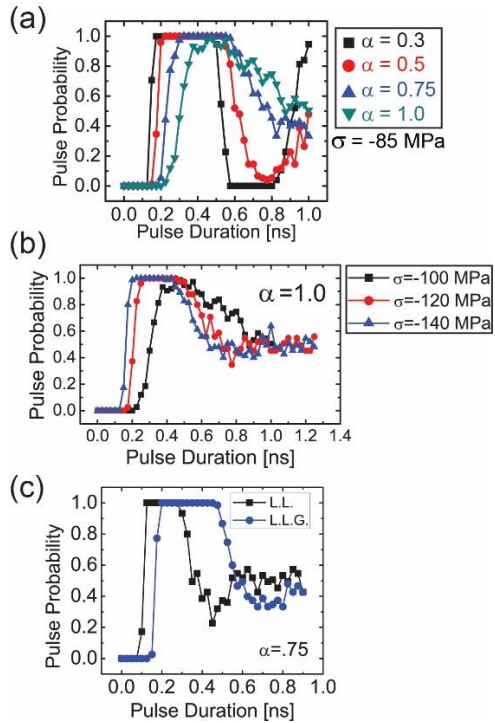

Figure 6. a) Dependence of the simulated pulse switching probability on $\alpha$ for $\sigma = -85$ MPa. b) Dependence of pulse switching probability on stress amplitude. Stress-induced switching is possible even for $\alpha = 1.0$. c) Comparison of pulse switching probability for LL and LLG dynamics for $\sigma = -85$ MPa and $\alpha = 0.75$. Here the difference between the LL and LLG dynamics has a significant effect on the width of the pulse window where reliable switching is predicted by the simulations ($\Delta \tau_{LL} = 200$ ps and $\Delta \tau_{LLG} = 320$ ps.)

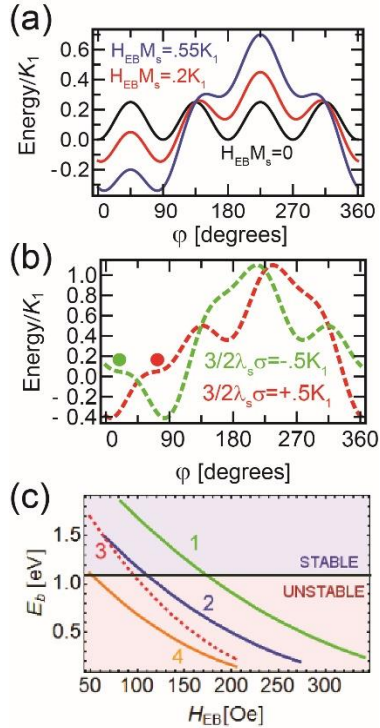

Figure 7. a) Energy (normalized to $K_1$) landscape as a function of angle for various values of exchange bias energy. b) $\varphi = 80°$ ($\varphi = 10°$) is the only stable equilibrium for compressive (tensile) stress. Dissipative dynamics and the free energy landscape then dictate the non-toggle switching dynamics. c) Shows the energy barrier dependence on the [110] bias field for a $d = 100$ nm, $t_{film} = 12.5$ nm circular element with (curve 1) $K_1 = 2.5 \times 10^5$ ergs/cm³, (curve 2) $K_1 = 2.0 \times 10^5$ ergs/cm³, and (curve 4) $K_1 = 1.5 \times 10^5$ ergs/cm³. Curve 3 shows the energy barrier dependence for $K_1 = 1.5 \times 10^5$ ergs/cm³ and $d = 100$ nm & $t_{film} = 15$ nm.

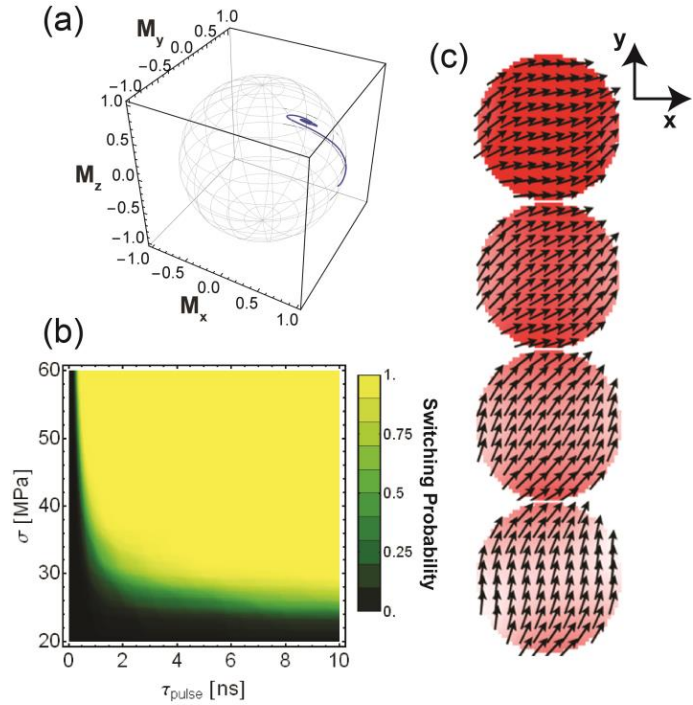

Figure 8. a) Magnetoelastic switching trajectory for $Fe_{81}Ga_{19}$ with $\sigma$ = -45 MPa and $\tau_{pulse}$ = 3 ns. The main part of the switching occurs within 200 ps. The magnetization relaxes to the equilibrium defined when the pulse is on and then relaxes to the final equilibrium when the pulse is turned off. b) Switching probability phase diagram for $Fe_{81}Ga_{19}$ with biaxial anisotropy at T = 300 °K. c) T = 0 °K OOMMF simulations showing the equilibrium micromagnetic configuration for $K_1 = 2\times10^5$ ergs/cm$^3$ and $M_s$ = 1300 emu/cm$^3$. Subsequent shots show the rotational switching mode for a 45 MPa uniaxial compressive stress along [100]. Color scale is blue-white-red indicating the local projection $m_x = -1$ (blue), $m_x = 0$ (white), $m_x = +1$ (red).

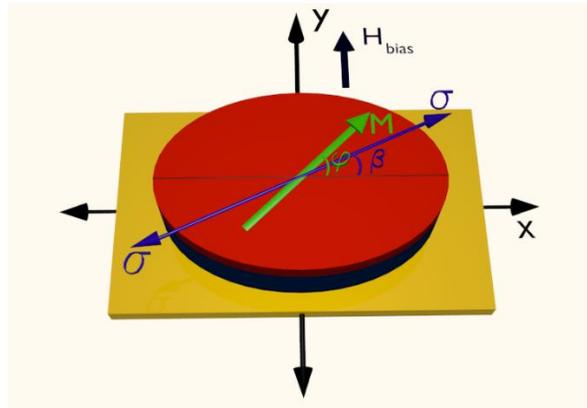

Figure 9. Schematic of magnetostrictive device geometry that utilizes uniaxial anisotropy to achieve deterministic switching. Polycrystalline $Tb_{0.3}Dy_{0.7}Fe_2$ on PMN-PT with 1 axis oriented at angle $\beta$ with respect to the easy axis. In this geometry, $\mathbf{M}$ lies in the x-y plane (film-plane) with the normalized $\mathbf{m} = \cos\varphi\hat{\mathbf{x}} + \sin\varphi\hat{\mathbf{y}}$.

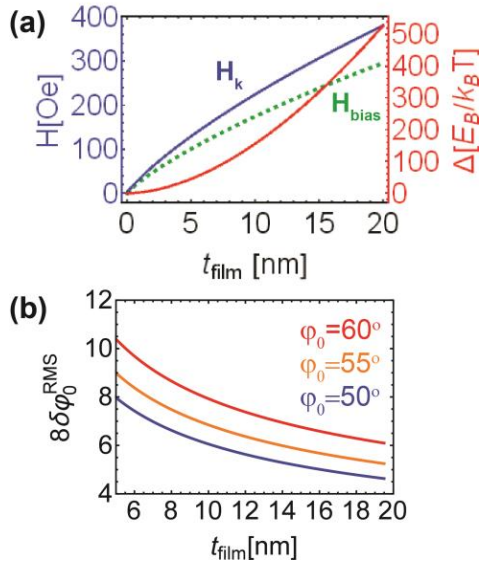

Figure 10. a) In-plane shape anisotropy field ($H_k$) and hard axis bias field ($H_{bias}$) for a 400×900 nm² ellipse as a function of film thickness required to ensure $\varphi_0 = 51°$. Thermal stability parameter $\Delta$ plotted versus film thickness with $H_k$, $H_{bias}$ such that $\varphi_0 = 51°$. b) Eight times the RMS angle fluctuation about three different average $\varphi_0 > 45°$ versus film thickness for a 400×900 nm² ellipse at T = 300 °K.

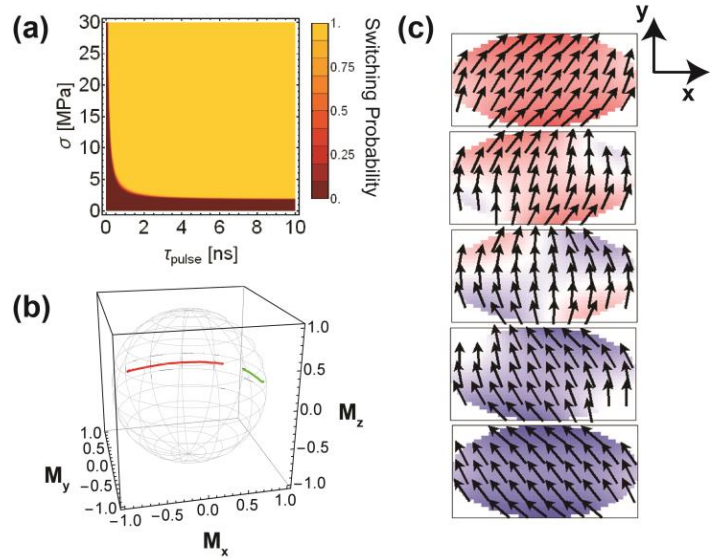

Figure 11. a) Magnetization trajectories for $\beta = 45°$, $\sigma = -5$ MPa, $\tau_{pulse} = 3$ ns, with $H_{bias} \sim 200$ Oe yielding $\varphi_0 = 51°$ (red) and $\beta = 45°, \sigma = -20$ MPa with $H_{bias} = 120$ Oe yielding $\varphi_0 = 28°$ (green). b) T = 300 °K stress pulse (compressive) switching probability phase diagram for a 400×900 nm² ellipse with $t_{film} = 12.5$ nm, $\beta = 45°$, $\varphi_0 = 51°$ c) Micromagnetic switching trajectory of a 400×900 nm² ellipse under a DC compressive stress of -3 MPa transduced along 45 degrees. Color scale is blue-white-red indicating the local projection $m_x = -1$ (blue), $m_x = 0$ (white), $m_x = +1$ (red).